# Comparative Analysis of 2D and 3D ResNet Architectures for IDH and MGMT Mutation Detection in Glioma Patients

**Danial Elyassirad[1], Benyamin Gheiji[1], Mahsa Vatanparast[1], Amir Mahmoud Ahmadzadeh, MD[2], Neda Kamandi, MD[3], Amirmohammad Soleimanian, MD[3], Sara Salehi, MD[4], Shahriar Faghani, MD[4*]**

(1) Student Research Committee, Mashhad University of Medical Sciences, Mashhad, Iran

(2) Department of Radiology, Mashhad University of Medical Sciences, Mashhad, Iran

(3) Faculty of Medicine, Mashhad University of Medical Sciences, Mashhad, Iran

(4) Radiology Informatics Lab, Department of Radiology, Mayo Clinic, Rochester, Minnesota

(*) Correspondence: Shahriar Faghani, Email: Faghani.Shahriar@mayo.edu

## Abstract

Gliomas are the most common cause of mortality among primary brain tumors. Molecular markers, including Isocitrate Dehydrogenase (IDH) and O[6]-methylguanine-DNA methyltransferase (MGMT) influence treatment responses and prognosis. Deep learning (DL) models may provide a non-invasive method for predicting the status of these molecular markers. To achieve non-invasive determination of gene mutations in glioma patients, we compare 2D and 3D ResNet models to predict IDH and MGMT status, using T1, post-contrast T1, and FLAIR MRI sequences. USCF glioma dataset was used, which contains 495 patients with known IDH and 410 patients with known MGMT status. The dataset was divided into training (60%), tuning (20%), and test (20%) subsets at the patient level. The 2D models take axial, coronal, and sagittal tumor slices as three separate models. To ensemble the 2D predictions the three different views were combined using logistic regression. Various ResNet architectures (ResNet10, 18, 34, 50, 101, 152) were trained. For the 3D approach, we incorporated the entire brain tumor volume in the ResNet10, 18, and 34 models. After optimizing each model, the models with the lowest tuning loss were selected for further evaluation on the separate test sets. The best-performing models in IDH prediction were the 2D ResNet50, achieving a test area under the receiver operating characteristic curve (AUROC) of 0.9096, and the 3D ResNet34, which reached a test AUROC of 0.8999. For MGMT status prediction, the 2D ResNet152 achieved a test AUROC of 0.6168; however, all 3D models yielded AUROCs less than 0.5. Overall, the study indicated that both 2D and 3D models showed high predictive value for IDH prediction, with slightly better performance in 2D models.

Keywords: Radiogenomics, Brain tumors, Molecular subtyping, Neural network, Machine learning

## ***Introduction***

Gliomas are highly heterogeneous, aggressive brain tumors that vary in terms of genetic mutations and histopathological features, significantly influencing prognosis and treatment strategies. The World Health Organization (WHO) classification of central nervous system tumors has undergone significant revisions in recent years particularly about incorporating molecular markers such as Isocitrate Dehydrogenase (IDH) mutation, which plays a crucial role in glioma stratification [1]. IDH mutations have become a central component of glioma classification due to their strong association with better prognosis and distinct clinical behavior compared to IDH-wildtype tumors [2,3]. In contrast, the O[6]-methylguanine-DNA methyltransferase (MGMT) promoter methylation status which reflects the tumor's ability to repair DNA damage, has emerged as an important predictive biomarker for glioma treatment, particularly in response to alkylating agents [4,5].

Clinical neuroimaging, combined with histomolecular evaluation of tumor samples, plays a crucial role in glioma diagnosis and treatment planning by providing comprehensive tissue information [6]. However, clinical images are often assessed qualitatively. The quantitative analysis of these images helps elucidate tumor biology and treatment response [7-10]. Radiogenomics, which integrates quantitative image analysis with genomic data, addresses these limitations [11,12]. Over the past decade, radiogenomics has shown significant promise in developing non-invasive prognostic and diagnostic tools, particularly for cancer, by linking imaging features with genomic signatures [13]. With advancements in molecular cancer characterization, texture analysis, and machine learning, radiogenomics supports personalized medicine, offering insights into diagnosis, treatment, prognosis, and optimal therapeutic strategies [14].

The recent WHO revision highlights the increasing importance of molecular markers in glioma diagnosis. Key changes include distinguishing between adult- and pediatric-type diffuse gliomas, introducing new diagnostic categories, refining the classification of IDH-mutant and IDH-wildtype gliomas, and incorporating molecular markers into tumor grading alongside traditional phenotypic assessments [1]. While molecular testing is considered the gold standard for identifying these markers it is not always accessible or feasible in all clinical settings. Radiogenomics offers a valuable alternative by using virtual biopsies to predict genetic mutations [15].

In recent years, ResNet models, which are a type of convolutional neural networks (CNNs), have gained popularity in medical imaging due to their high performance with relatively lower computational demands. Several studies have employed ResNet for radiogenomics, however, despite the frequent use of ResNet models in radiogenomics, there has been little to no comparison between 2D and 3D variants of these models. In this study, we used different 2D and 3D ResNet models to predict IDH and MGMT mutations in glioma patients and systematically compared the challenges and outcomes associated with each approach.

## ***Method***

This study used pre-operative T1-weighted (T1), T1-weighted contrast-enhanced (T1c), and fluid-attenuated inversion recovery (FLAIR) MRI sequences from 495 patients diagnosed with CNS gliomas of biopsy proven WHO grades 2-4 from UCSF glioma dataset [16]. Preoperative MRI scans were obtained using a 3.0 Tesla scanner and a dedicated 8-channel head coil. Inclusion criteria for the study required the availability of T1, T1c, and FLAIR MRI sequences. The dataset included 495 patients with known IDH status, with 103 identified as IDH-mutated and 392 as wild-type. Furthermore, 410 patients were evaluated for MGMT promoter methylation status, with 297 positive and 113 negative cases based on an MGMT index ranging from 0-17. An MGMT index of 0 was defined as negative, while any index above 0 was considered positive. The dataset was subsequently divided into three subsets at the patient-level: 60% for training, 20% for tuning, and 20% for testing [17].

In this study, we used the dataset's available segmentations. Each segmentation mask consisted of three subregions including enhancing tumor, non-enhancing/necrotic tumor, and surrounding FLAIR abnormality areas. For both 2D and 3D approaches, the segmentation masks were overlaid onto the MR images and the area of the masks were extracted from original images to be used as deep learning (DL) models input. These processed images were normalized and scaled to their respective intensity ranges to maintain consistency. An intensity range was then allocated to each subregion of the segmentation masks to better represent the differences of subregions' intensity.

For the 2D approach, axial, coronal, and sagittal masked slices containing the largest area of tumor regions —identified by the counts of non-zero voxels of mask— were selected. We used three single-channel ResNet models (including ResNet10, 18, 34, 50, 101, 152), each trained on individual views, and employed an ensemble method to combine their predictions using logistic regression (LR). In the 3D approach, entire brain tumor volumes were used as the inputs of the ResNet10, 18, and 34 models to generate a prediction.

For both 2D and 3D approaches, we employed a highly structured data processing pipeline, utilizing the medical imaging framework MONAI to ensure robust handling and analysis of the data [18]. We utilized a comprehensive suite of aggressive on-the-fly data augmentation strategies, such as flipping, rotations, zooming, intensity shifting, scaling, gaussian noise addition, contrast adjustment, gaussian smoothing, elastic deformation, grid distortion, and histogram shifting. These techniques aimed to enhance the model's ability to generalize across various imaging characteristics, rather than to balance data among classes. The model was trained using a binary cross-entropy loss function that was weighted to tackle the inherent class imbalances within the dataset [19].

Dropout, early stopping, learning rate reduction, and model checkpoint callbacks were utilized to further optimize training. Custom data generators were used for batching, ensuring efficient feeding of images into the model. The best models, selected based on the lowest tuning loss, were used to predict IDH and MGMT mutations on the test set. The performance was evaluated using the area under the receiver operating characteristic curve (AUROC) on the tuning and test set [20]. Also, to compare the performance of different 2D and 3D ResNet architectures on the test set, we plotted the test AUROC along with the regression line.

## ***Results***

Table 1 presents a summary of the demographic and clinical data for the patients involved in this study.

| | | All patients | IDH | | MGMT | |
|---|---|---|---|---|---|---|
| | | | Mutated | Wild-type | Positive | Negative |
| Number of patients | | 495 | 103 | 392 | 297 | 113 |
| Age | | 56.87 (17-94) | 38.81 (17-71) | 61.61 (21-94) | 59.32 (19-94) | 59.81 (17-89) |
| Sex | Female | 199 | 40 | 159 | 124 | 36 |
| | Male | 296 | 63 | 233 | 173 | 77 |
| IDH | Mutated | 103 | | | 37 | 4 |
| | Wildtype | 392 | | | 260 | 109 |
| MGMT status | Positive | 297 | 37 | 260 | | |
| | Negative | 113 | 4 | 109 | | |
| MGMT index | | 6.76 (0-17) | 10.48 (0-17) | 6.36 (0-17) | 9.33 (1-16) | 0 |
| Final pathologic diagnosis | Astrocytoma | 114 | 90 | 24 | 49 | 9 |
| | Glioblastoma | 368 | 0 | 368 | 247 | 104 |
| | Oligodendroglioma | 13 | 13 | 0 | 1 | 0 |
| WHO CNS Grade | 2 | 56 | 46 | 10 | 7 | 1 |
| | 3 | 43 | 29 | 14 | 16 | 7 |
| | 4 | 396 | 28 | 368 | 274 | 105 |
| Survival | Dead | 248 | 11 | 237 | 163 | 69 |
| | Alive | 247 | 92 | 155 | 134 | 44 |
| Overall survival | | 573.87 | 973.42 | 468.62 | 534.17 | 463.73 |
| Maximizing extent of resection | Biopsy | 54 | 4 | 50 | 29 | 12 |
| | Gross total resection | 244 | 28 | 216 | 163 | 61 |
| | Subtotal resection | 196 | 71 | 125 | 104 | 40 |

Table 1. Demographic and clinical data of the patients

The mean and range of AUROCs for 2D models in axial, coronal, and sagittal view across different sequences were 0.8444 (range: 0.6520 - 0.9017), 0.8344 (range: 0.7271 - 0.8767), and 0.8718 (range: 0.8211 - 0.8968). Also, mean and range AUROCs for 2D ensemble models across all sequences was 0.8782 (range: 0.8199-0.9096). For the 3D models, the mean AUROCs across all sequences was 0.8586 (range: 0.8272-0.8999). Regarding both methods (ensemble 2D models and 3D models), the mean test AUROC of all models was 0.8717 (range: 0.8199-0.9096).

In both 2D and 3D models, based on the mean test AUROC, T1c had superior performance with mean test AUROC of 0.8924 in 2D models and 0.8881 in 3D models. 2D models reached mean test AUROC of 0.8883 and 0.8540 in T1 and FLAIR sequences, and for 3D models, these AUROCs were 0.8490 and 0.8388 for T1 and FLAIR sequences, respectively.

Among all models, the best-performing for IDH detection on the test set were ResNet50 in 2D models, with a test AUROC of 0.9096 using the T1, and ResNet34 for the 3D models, which reached a test AUROC of 0.8999 using the T1c.

The performance of the 2D and 3D ResNet models was reported in Tables 2-3. Figures 1-2 display the models performance for IDH predictions across different sequences.

| Model | Gene | Modality | View | Val AUROC | Test AUROC | Ensemble test AUROC | Modality | View | Val AUROC | Test AUROC | Ensemble test AUC | Modality | View | Val AUROC | Test AUROC | Ensemble test AUROC |
|---|---|---|---|---|---|---|---|---|---|---|---|---|---|---|---|---|
| | | | Axial | 0.8736 | 0.8474 | | | Axial | 0.9139 | 0.8687 | | | Axial | 0.6624 | 0.6520 | |
| Resnet10, 2D | IDH | T1 | Coronal | 0.8858 | 0.8455 | 0.8736 | T1c | Coronal | 0.9164 | 0.8767 | 0.8913 | FLAIR | Coronal | 0.8840 | 0.8059 | 0.8199 |
| | | | Sagittal | 0.9023 | 0.8761 | | | Sagittal | 0.9078 | 0.8846 | | | Sagittal | 0.8529 | 0.8211 | |
| | | | Axial | 0.9017 | 0.8669 | | | Axial | 0.9182 | 0.8761 | | | Axial | 0.8950 | 0.8236 | |
| Resnet18, 2D | IDH | T1 | Coronal | 0.8742 | 0.8455 | 0.8730 | T1c | Coronal | 0.9158 | 0.8608 | 0.8962 | FLAIR | Coronal | 0.9072 | 0.8407 | 0.8516 |
| | | | Sagittal | 0.9060 | 0.8681 | | | Sagittal | 0.9249 | 0.8919 | | | Sagittal | 0.8938 | 0.8541 | |
| | | | Axial | 0.9054 | 0.8767 | | | Axial | 0.9212 | 0.8724 | | | Axial | 0.8907 | 0.8266 | |
| Resnet34, 2D | IDH | T1 | Coronal | 0.8864 | 0.8425 | 0.8956 | T1c | Coronal | 0.9072 | 0.8669 | 0.8822 | FLAIR | Coronal | 0.7937 | 0.7271 | 0.8602 |
| | | | Sagittal | 0.9151 | 0.8626 | | | Sagittal | 0.9158 | 0.8755 | | | Sagittal | 0.9103 | 0.8578 | |
| | | | Axial | 0.9096 | 0.9017 | | | Axial | 0.9176 | 0.8761 | | | Axial | 0.8767 | 0.8016 | |
| Resnet50, 2D | IDH | T1 | Coronal | 0.8834 | 0.8309 | 0.9096 | T1c | Coronal | 0.9072 | 0.8718 | 0.8956 | FLAIR | Coronal | 0.8669 | 0.8223 | 0.8700 |
| | | | Sagittal | 0.9072 | 0.8828 | | | Sagittal | 0.9335 | 0.8773 | | | Sagittal | 0.9084 | 0.8761 | |
| | | | Axial | 0.8901 | 0.8565 | | | Axial | 0.9316 | 0.8816 | | | Axial | 0.9103 | 0.8455 | |
| Resnet101, 2D | IDH | T1 | Coronal | 0.8803 | 0.8248 | 0.8987 | T1c | Coronal | 0.9109 | 0.8675 | 0.8956 | FLAIR | Coronal | 0.8779 | 0.7729 | 0.8718 |
| | | | Sagittal | 0.9090 | 0.8871 | | | Sagittal | 0.9096 | 0.8968 | | | Sagittal | 0.9115 | 0.8571 | |
| | | | Axial | 0.9176 | 0.8413 | | | Axial | 0.9328 | 0.8736 | | | Axial | 0.8498 | 0.8107 | |
| Resnet152, 2D | IDH | T1 | Coronal | 0.8846 | 0.8455 | 0.8791 | T1c | Coronal | 0.9164 | 0.8639 | 0.8932 | FLAIR | Coronal | 0.9035 | 0.8077 | 0.8504 |
| | | | Sagittal | 0.9255 | 0.8736 | | | Sagittal | 0.9249 | 0.8852 | | | Sagittal | 0.9115 | 0.8645 | |

Table 2. 2D models performances for IDH mutation classification

| Model | Gene | Modality | Val AUROC | Test AUROC | Modality | Val AUROC | Test AUROC | Modality | Val AUROC | Test AUROC |
|---|---|---|---|---|---|---|---|---|---|---|
| Resnet10, 3D | IDH | T1 | 0.8864 | 0.8272 | T1c | 0.9396 | 0.8700 | FLAIR | 0.9237 | 0.8278 |
| Resnet18, 3D | IDH | T1 | 0.9237 | 0.8443 | T1c | 0.9341 | 0.8944 | FLAIR | 0.9164 | 0.8309 |
| Resnet34, 3D | IDH | T1 | 0.8889 | 0.8755 | T1c | 0.9560 | 0.8999 | FLAIR | 0.9188 | 0.8578 |

Table 3. 3D models performances for IDH mutation classification

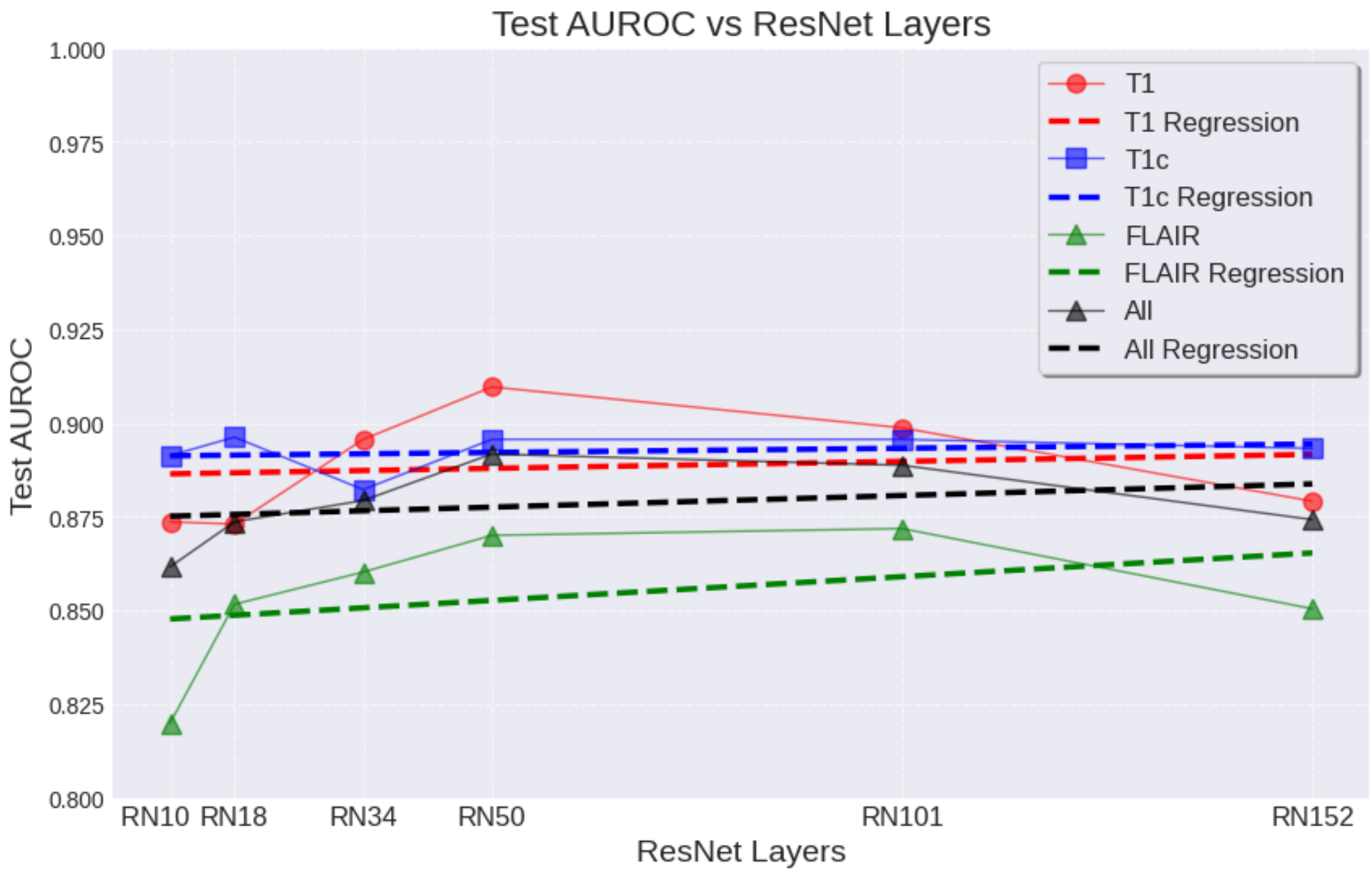

Figure 1. Test AUROC of 2D models in IDH prediction

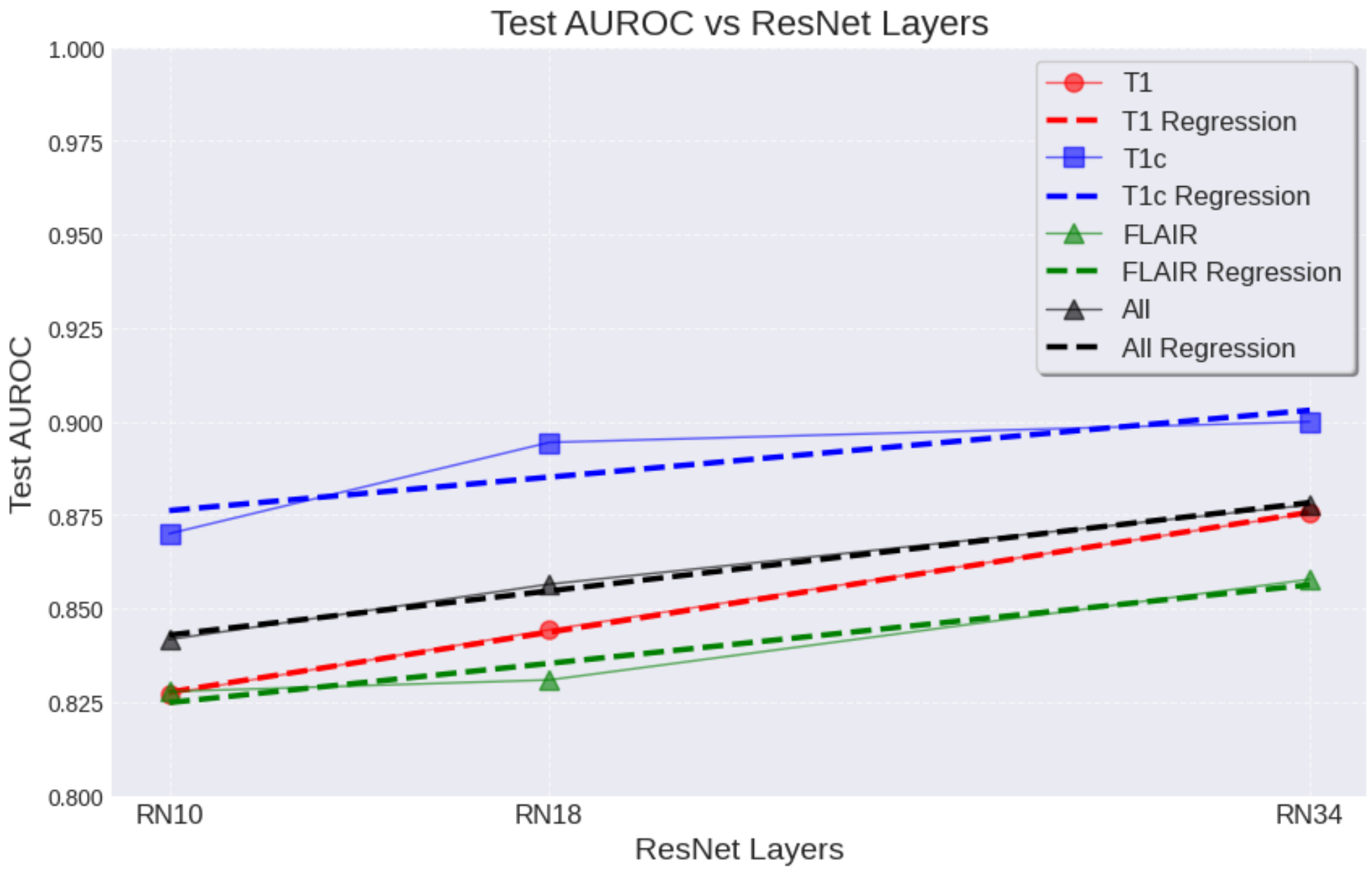

Figure 2. Test AUROC of 3D models in IDH prediction

In contrast, for MGMT mutation detection, the test AUROCs were significantly lower across all models. The best-performing model for determining MGMT mutation status, the 2D ResNet152 with T1, achieved a test AUROC of 0.6168. Other models and sequences yielded lower AUROCs, with several cases showing AUROCs below 0.5 (14 out of 16 models).

## ***Discussion***

In this study, the performance of 2D and 3D ResNet models in detecting IDH and MGMT mutations using MRI was assessed. For IDH prediction, the overall mean test AUROC of 2D models was 0.8782 compared to 0.8586 for 3D models. In contrast, MGMT mutation detection results were significantly lower, with the best model, the 2D ResNet152 using T1c, yielding a test AUROC of 0.6168.

In this study, we used various 2D and 3D ResNet architectures. We utilized ResNet due to several reasons: Firstly, ResNet is widely used in medical imaging. Secondly, ResNet's architecture demands fewer resources compared to more sophisticated models, a significant advantage given the resource constraints common in DL models.

To predict IDH and MGMT mutations in glioma patients using MRI, we faced several challenges. The primary challenge was the limited availability of publicly available data, a common issue in medical imaging DL. Several strategies can help mitigate this challenge, including data augmentation, leveraging pre-trained models, and synthetic data generation. We chose data augmentation for several reasons: it is more resource-efficient and increases model flexibility. We opted against using pre-trained models due to the limited availability of 3D pre-trained networks, many of which rely on 2D weights. While synthetic data generation could potentially expand the dataset, it is a resource intensive approach.

When using 2D models, potential discriminating imaging features might be omitted. To partially address this, we employed an ensemble strategy, using three different models trained on the largest tumor slice from the axial, sagittal, and coronal views. This approach allowed us to retain more information.

For the IDH determination task, the highest test AUROC was achieved by a 2D model, specifically ResNet50 using the T1, with a test AUROC of 0.9096. In comparison, the best-performing 3D model, ResNet34 with the T1c, achieved a test AUROCs of 0.8999. The mean test AUROC for 2D and 3D models were 0.8782 and 0.8586, respectively.

There are no notable differences between the performance of 2D and 3D models, suggesting that 3D models are not always superior to their 2D counterparts. This may be attributed to the fact that using the largest masked slice across three views provides substantial information about the tumor, offering valuable data for accurate IDH prediction.

As shown in Figures 1 and 2, more complex models generally yield better results in both 2D and 3D models, with linear regression lines for each sequence showing positive slopes, especially in the 3D models. This suggests that more complex ResNets may be better at identifying effective features for predicting IDH mutations. Furthermore, in the 3D models, ResNets with more layers consistently demonstrated better test AUROCs than those with fewer layers within the same sequence.

Given that IDH status is predictable, as demonstrated by our study and numerous others, future research should focus on developing more reliable models with larger datasets and incorporating external validation [21-23]. By incorporating local datasets from various regions, these steps could greatly enhance the clinical applicability of models for classifying glioma patients, especially those with tumors located in high-risk areas that are difficult to biopsy. One of the most critical aspects moving forward is the implementation of uncertainty quantification, which provides confidence level for prediction, leading to more dependable results on a case-by-case basis [24].

We aimed to explore the prediction of MGMT promoter methylation status in glioma patients using MRI, a task that remains both challenging and contentious. While several studies have demonstrated promising results, the inconsistencies across datasets and methodologies highlight significant obstacles. For instance, Chang et al. achieved 83% accuracy using a ResNet-based approach on multimodal MRI data, employing data from The Cancer Imaging Archives (TCIA) and The Cancer Genome Atlas (TCGA) [25]. Radiomics-based methods, such as those proposed by Le et al. and Do et al., also reported impressive accuracies of 89% and 93%, respectively, by leveraging advanced feature selection techniques like F-scores and genetic algorithms [26,27]. These results suggest that, under controlled conditions, MRI might provide valuable insights into MGMT promoter methylation.

However, contrasting findings have emerged from the RSNA-MICCAI brain tumor radiogenomic classification competition, where the best model only achieved 62% accuracy on the BraTS dataset. This disparity could be attributed to several factors, as highlighted by Faghani et al. One potential explanation for the poorer results on the BraTS dataset is the heterogeneity in labeling methods across contributing institutions. The MGMT promoter methylation status was determined using various techniques, introducing noisy labels and making consistent prediction difficult. Additionally, the multi-institutional nature of the BraTS dataset may result in a more realistic yet challenging environment for model development, reflecting the complexity of real-world clinical scenarios [28].

In our study, we achieved an AUROC of 0.6168 on a separate test set for MGMT using 2D models. However, for the 2D models, 14 out of 16 final models reached an AUROC of less than 0.5. Furthermore, Saeed et al. reported that despite developing diverse DL models, their results barely exceeded random chance, reinforcing the notion that there may be no clear correlation between imaging data and MGMT methylation status [29].

Similarly, Robinet et al. provided insights into why some studies have yielded better results. Factors such as small sample sizes, potential data leakage, and the absence of independent test sets for validation have contributed to inflated performance metrics in some studies [30].

In conclusion the task of predicting MGMT promoter methylation from MRI scans remains unresolved. While some models report high accuracy, issues such as data heterogeneity, potential data leakage, small sample sizes, and lack of reproducibility significantly hinder progress. As Robinet et al. pointed out, it may not be possible with current algorithms to extract reliable information about the MGMT biomarker from MRI. We believe MGMT status poor prediction results may be related to two factors. First, MGMT promoter methylation status may not significantly influence tumor imaging characteristics. Second, the cutoff used to define negative and positive MGMT status labels may be problematic. In the UCSF dataset, MGMT status of 0 is labeled as negative, and values between 1 and 17 are considered positive. However, this cutoff seems suboptimal. A standardized cutoff for MGMT status should be established, but it is also important to recognize that MGMT status may have only a limited impact on imaging tumor characteristics.

This study demonstrates that 2D ensemble models, employing three different views, can predict IDH mutation status as effectively as 3D models, without notable differences in performance but with reduced resource utilization. Additionally, we found a correlation between the complexity of models and their performance on the test set, particularly notable in 3D models. Despite these findings, the study recognizes several limitations, including limited data and the absence of external validation. Also for future studies, we recommend incorporating uncertainty quantification for radiogenomics prediction and expanding the dataset size to improve model robustness. Additionally, for MGMT prediction, a standardized MGMT assay, which includes using a consistent method for calculating MGMT status across all patients and implementing an optimized threshold could be beneficial. These improvements will enhance the reliability and clinical applicability of the radiogenomics models.

## ***Declarations***

AI assistant tools

We used ChatGPT-4 to enhance grammar and manuscript content. All the authors reviewed and edited the content to avoid plagiarism.

Acknowledgements

The authors declare that no funds, grants, conflict of interest, or other support were received during the preparation of this manuscript.

## ***References***

1. Louis DN, Perry A, Wesseling P, Brat DJ, Cree IA, Figarella-Branger D, Hawkins C, Ng HK, Pfister SM, Reifenberger G, Soffietti R. The 2021 WHO classification of tumors of the central nervous system: a summary. Neuro-oncology. 2021 Aug 1;23(8):1231-51.

2. Yan H, Parsons DW, Jin G, McLendon R, Rasheed BA, Yuan W, Kos I, Batinic-Haberle I, Jones S, Riggins GJ, Friedman H. IDH1 and IDH2 mutations in gliomas. New England journal of medicine. 2009 Feb 19;360(8):765-73.

3. Sun H, Yin L, Li S, Han S, Song G, Liu N, Yan C. Prognostic significance of IDH mutation in adult low-grade gliomas: a meta-analysis. Journal of Neuro-oncology. 2013 Jun;113:277-84.

4. Shaw R, Basu M, Karmakar S, Ghosh MK. MGMT in TMZ-based glioma therapy: multifaceted insights and clinical trial perspectives. Biochimica et Biophysica Acta (BBA)-Molecular Cell Research. 2024 Jan 18:119673.

5. Reifenberger G, Wirsching HG, Knobbe-Thomsen CB, Weller M. Advances in the molecular genetics of gliomas—implications for classification and therapy. Nature reviews Clinical oncology. 2017 Jul;14(7):434-52.

6. Lohmann P, Galldiks N, Kocher M, Heinzel A, Filss CP, Stegmayr C, Mottaghy FM, Fink GR, Shah NJ, Langen KJ. Radiomics in neuro-oncology: Basics, workflow, and applications. Methods. 2021 Apr 1;188:112-21.

7. Tomaszewski MR, Gillies RJ. The biological meaning of radiomic features. Radiology. 2021 Mar;298(3):505-16.

8. Fathi Kazerooni A, Bakas S, Saligheh Rad H, Davatzikos C. Imaging signatures of glioblastoma molecular characteristics: a radiogenomics review. Journal of Magnetic Resonance Imaging. 2020 Jul;52(1):54-69.

9. Gillies RJ, Kinahan PE, Hricak H. Radiomics: images are more than pictures, they are data. Radiology. 2016 Feb;278(2):563-77.

10. Gillies RJ, Schabath MB. Radiomics improves cancer screening and early detection. Cancer Epidemiology, Biomarkers & Prevention. 2020 Dec 1;29(12):2556-67.

11. Liu Z, Duan T, Zhang Y, Weng S, Xu H, Ren Y, Zhang Z, Han X. Radiogenomics: a key component of precision cancer medicine. British Journal of Cancer. 2023 Sep 21;129(5):741-53.

12. Song L, Zhu Z, Mao L, Li X, Han W, Du H, Wu H, Song W, Jin Z. Clinical, conventional CT and radiomic feature-based machine learning models for predicting ALK rearrangement status in lung adenocarcinoma patients. Frontiers in Oncology. 2020 Mar 20;10:369.

13. Moassefi M, Faghani S, Erickson BJ. Artificial Intelligence in Neuro-Oncology: predicting molecular markers and response to therapy. Medical Research Archives. 2024 Jun 24;12(6).

14. Shui L, Ren H, Yang X, Li J, Chen Z, Yi C, Zhu H, Shui P. The era of radiogenomics in precision medicine: an emerging approach to support diagnosis, treatment decisions, and prognostication in oncology. Frontiers in Oncology. 2021 Jan 26;10:570465.

15. Moassefi M, Erickson BJ. Bridging Pixels to Genes. Radiology: Artificial Intelligence. 2024 Jun 20;6(4):e240262.

16. Calabrese E, Villanueva-Meyer JE, Rudie JD, Rauschecker AM, Baid U, Bakas S, Cha S, Mongan JT, Hess CP. The University of California San Francisco preoperative diffuse glioma MRI dataset. Radiology: Artificial Intelligence. 2022 Oct 5;4(6):e220058.

17. Rouzrokh P, Khosravi B, Faghani S, Moassefi M, Vera Garcia DV, Singh Y, Zhang K, Conte GM, Erickson BJ. Mitigating bias in radiology machine learning: 1. Data handling. Radiology: Artificial Intelligence. 2022 Aug 24;4(5):e210290.

18. Cardoso MJ, Li W, Brown R, Ma N, Kerfoot E, Wang Y, Murrey B, Myronenko A, Zhao C, Yang D, Nath V. Monai: An open-source framework for deep learning in healthcare. arXiv preprint arXiv:2211.02701. 2022 Nov 4.

19. Zhang K, Khosravi B, Vahdati S, Faghani S, Nugen F, Rassoulinejad-Mousavi SM, Moassefi M, Jagtap JM, Singh Y, Rouzrokh P, Erickson BJ. Mitigating bias in radiology machine learning: 2. Model development. Radiology: Artificial Intelligence. 2022 Aug 24;4(5):e220010.

20. Faghani S, Khosravi B, Zhang K, Moassefi M, Jagtap JM, Nugen F, Vahdati S, Kuanar SP, Rassoulinejad-Mousavi SM, Singh Y, Vera Garcia DV. Mitigating bias in radiology machine learning: 3. Performance metrics. Radiology: Artificial Intelligence. 2022 Aug 24;4(5):e220061.

21. Wu X, Zhang S, Zhang Z, He Z, Xu Z, Wang W, Jin Z, You J, Guo Y, Zhang L, Huang W. Biologically interpretable multi-task deep learning pipeline predicts molecular alterations, grade, and prognosis in glioma patients. NPJ Precision Oncology. 2024 Aug 16;8(1):181.

22. McHugh H, Safaei S, Maso Talou GD, Gock SL, Yeun Kim J, Wang A. IDH and 1p19q Diagnosis in Diffuse Glioma from Preoperative MRI Using Artificial Intelligence. medRxiv. 2023 Apr 29:2023-04.

23. Zhong S, Ren JX, Yu ZP, Peng YD, Yu CW, Deng D, Xie Y, He ZQ, Duan H, Wu B, Li H. Predicting glioblastoma molecular subtypes and prognosis with a multimodal model integrating convolutional neural network, radiomics, and semantics. Journal of Neurosurgery. 2022 Dec 2;139(2):305-14.

24. Faghani S, Moassefi M, Rouzrokh P, Khosravi B, Baffour FI, Ringler MD, Erickson BJ. Quantifying uncertainty in deep learning of radiologic images. Radiology. 2023 Aug 1;308(2):e222217.

25. Chang P, Grinband J, Weinberg BD, Bardis M, Khy M, Cadena G, Su MY, Cha S, Filippi CG, Bota D, Baldi P. Deep-learning convolutional neural networks accurately classify genetic mutations in gliomas. American Journal of Neuroradiology. 2018 Jul 1;39(7):1201-7.

26. Le NQ, Do DT, Chiu FY, Yapp EK, Yeh HY, Chen CY. XGBoost improves classification of MGMT promoter methylation status in IDH1 wildtype glioblastoma. Journal of personalized medicine. 2020 Sep 15;10(3):128.

27. Do DT, Yang MR, Lam LH, Le NQ, Wu YW. Improving MGMT methylation status prediction of glioblastoma through optimizing radiomics features using genetic algorithm-based machine learning approach. Scientific Reports. 2022 Aug 4;12(1):13412.

28. Faghani S, Khosravi B, Moassefi M, Conte GM, Erickson BJ. A comparison of three different deep learning-based models to predict the MGMT promoter methylation status in glioblastoma using brain MRI. Journal of Digital Imaging. 2023 Jun;36(3):837-46.

29. Saeed N, Ridzuan M, Alasmawi H, Sobirov I, Yaqub M. MGMT promoter methylation status prediction using MRI scans? An extensive experimental evaluation of deep learning models. Medical Image Analysis. 2023 Dec 1;90:102989.

30. Robinet L, Siegfried A, Roques M, Berjaoui A, Cohen-Jonathan Moyal E. MRI-based deep learning tools for mgmt promoter methylation detection: a thorough evaluation. Cancers. 2023 Apr 12;15(8):2253.